\documentclass{article}
\usepackage{fleqn}
\usepackage{epsfig}
\usepackage{amsmath,amssymb}
\newcommand{\Fig}[4]{%
\begin{center}
\parbox{#2cm}{%
\refstepcounter{figure}\includegraphics[width=#2cm,
height=#3cm]{#1}}\\[12pt] \parbox{14cm}{\noindent {\bf Fig. \thefigure.}\quad
#4}\end{center}}

\newcommand{\TwoFigs}[5]{%
\begin{flushleft}
\parbox{158mm}{%
\begin{tabular}{ll}
\includegraphics[width=7.5cm,height=#5cm]{#1} & \includegraphics[width=7.5cm,height=#5cm]{#2}\\
\parbox{7.5cm}{{%
\refstepcounter{figure}\bf Fig. \thefigure.}\quad #3} & \parbox{7.5cm}{{%
\refstepcounter{figure}\bf Fig. \thefigure.}\quad #4} \\
\end{tabular}}
\end{flushleft}}

\begin{document}

\begin{center}
{\bf\Large A Mathematical Modeling In Computer Algebra Systems (CAS) As A Base For A Development of The Mathematics Education}\\[12pt]
Yurii Ignat'ev and Alsu Samigullina\\
Kazan Federal University,\\ Kremlyovskaya str., 35,
Kazan 420008, Russia
\end{center}

{\bf keywords}: IT; Physical and Mathematical Education; Mathematical Modeling; Computer Algebra Systems; Actual Training Technologies; Score-Rating System.\\
{\bf PACS}: 	01.40.gb , 01.50.H, 02.10.Ud, 02.10.Yn,	02.40.Dr, 02.50.Cw  

\begin{abstract}
Information technologies for studying physical-mathematical disciplines on base of mathematical modeling in the computer algebra system Maple are described.\end{abstract}

\section{The Necessity of Implementing Information Technolgies Into the Structure of Physical - Mathematical Education}

There exist a series of reasoned causes of the necessity to implement information technologies into the structure of physical-mathematical education. Basically, these reasons possess a character external with respect to physical-mathematical education and are inducted by global changes in the social structure, public conscience, and intensive process of IT penetration into the society. Let us list some of these reasons:

\begin{enumerate}
\item  information flows which are increasing continuously and rapidly along with quick obsolescence of the information;
\item  shortage of academic curricular hours dedicated to the study of fundamental disciplines with simultaneous  extension of the domains studied;
\item  change of the education process focus to the independent work of students and scholar alumni;
\item  the lack of funding of fundamental directions of science and corresponding directions of the higher education;
\item  integration of new domains of knowledge and appearance of new directions of science and technologies;
\item  increase of specialties' number with simultaneous decrease of the number of students;
\item  decline in interest of young people to science-driven specialties of physical-mathematical profiles since these require more efforts for their acquirement while the expected carrier growth does not give too large hopes for the future.
\item  Decline in the level of mathematical preparation of university entrants, including the result of factors of the previous item.
\end{enumerate}
Increase of the mass media and visual parts of the information (TV, Internet, videogames etc.) resulting in the violation of the analytical and abstract thinking which are necessary for the mathematical education. However, in addition to the above-cited causes, there exist, by our opinion,  some internal causes of namely Russian mathematical education which in recent times  start to lead towards its both stagnation and lower efficiency. Among these reasons we can indicate:
\begin{enumerate}
\item  formalized character of mathematical education;
\item  loss of relations of mathematical education with the contemporary problems of both fundamental and applied sciences;
\item  overload of mathematical disciplines by abstract theoretical materials to the detriment of solving concrete problems which historically are core ones in such courses;
\item  a gap between mathematical courses and contemporaneous computer technologies.
\end{enumerate}

Similar problems are peculiar particularly to many modern Russian mathematical schools. There are known, for example, the needs of many domains of both fundamental and applied sciences in creation of methods of investigation of nonlinear continual systems described by nonlinear differential and integral-differential partial differential equations. However, the majority of candidate of science and doctor of science thesises in this specialty are dedicated to the methods for solving linear differential and integral equations; frequently, these investigations are concluded with a proof of existence and uniqueness of solution.

Similar problems do exist in other countries, which are traditionally oriented towards science-driven forms of human activities. In last time, this situation creates prerequisites for stagnation of progressive development of the mankind. Therefore, in spite of the fact that in the present paper we mainly base upon the studies and experience of Russian scientists, the results exposed in the article would be useful for mathematicians of other countries who work in higher education.

\section{Basic Concept Of Implementation of the IT In Physical - Mathematical Education}

We should note that up to now peculiarities of the computerization of mathematical education on the base of modern information technologies are poorly elucidated in the scientific literature. In particular, there fails to exist a unified concept of the IT implementation into the structure of mathematical education; there is also lack of sufficiently developed education methods of teaching mathematical disciplines using information technologies with the specificity of these disciplines taken  into account. Existing methods are weakly related to the specificity of physical-mathematical disciplines allowing to realize  more deep penetration of information technologies to the essence of these subjects and permitting to reorient essentially the educative process, making it thus more efficient.

From our point of view the overcome of the mentioned contradictions between the demands of modern science and technology on the one hand and the potential of the mathematical education on the other hand, is possible by following the way of intense application of the mathematical and computer modeling during the study of all the basic mathematical courses with a following integration of these courses' target problems with problems of fundamental and applied sciences . In addition, the computer modeling must be carried out in the media of computer algebra systems (CAS), while the respective education tasks should be formatted as research ones appointed to construction of mathematical and computer models. During creation of these models, students will acquire necessary fundamental knowledge of subjects and will learn how to apply them in practice. It is noteworthy that both the construction of a mathematical model and its posterior computer realization train the exactitude of mathematical thinking as well as its culture and technological efficiency.  The construction and study of mathematical model train, in addition to other features, the diligence, accuracy, and fairness, i.e., the qualities which are seemingly lacking in post-soviet generations of youth. On the top of everything else, this way is the most efficient method to involve the young people in the modern science and engineering.

Academician A.A. Samarsky \cite{Samar}, one of the creators of the scientific direction of mathematical modeling, determined a mathematical model as an equivalent of an object, which reflects in a mathematical form its most important properties. He introduced the concept of the \textit{triad of mathematical modeling }(see Fig. 1): model > algorithm > program as the necessary plan of actions in studying the object.  Herewith, at the first stage, a mathematical image of an object is constructed, reflecting in mathematical form its most important properties, i.e., a \textit{mathematical model}.

\Fig{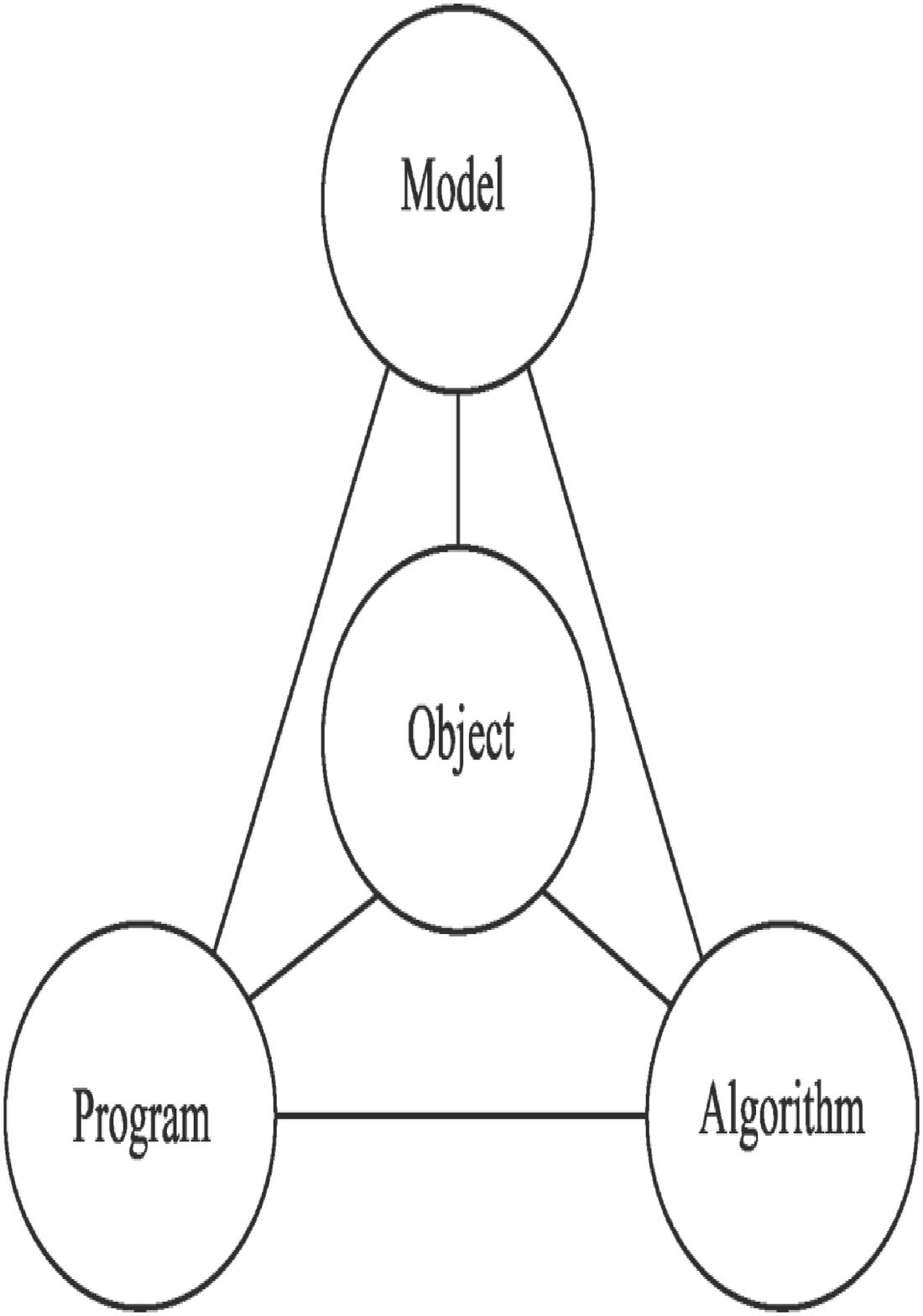}{8}{6}{Triad of mathematical modeling by Samarsky}

The mathematical model is further studied by theoretical methods, which makes it possible to obtain \textit{general preliminary }knowledge about the object. On the second stage, an \textit{algorithm }is developed for realization of the model using the computer. At this stage, the model is represented in a form convenient for applying numerical methods and a sequence of computational and logical operations for the study of investigated properties of the object is determined. At the third stage, which is the \textit{programming}, programs translating the model and algorithm into a program language are created. Having created the indicated triad of mathematical modeling, the researcher carries out \textit{numerical experiments}; comparing the results of which to those obtained via natural experiments, he/she then introduces the necessary corrections into the mathematical model. Realizing that way the refinement of the mathematical model up to an ideal state, the researcher thereby obtains the mathematical model adequate to the object.  Let us note that the proper process of construction of the mathematical model itself reflects adequately the process of outer world insight by a man. Therefore, this proves that the process of model construction ideally suits as a base for the construction of the model of mathematical education informatization.  By our opinion, namely this triad of mathematical modeling should be set into the foundation of the mathematical education.

The requirement of informatization of mathematical education leads to the necessity of deep implementation of the IT leveraging the mathematical and computer modeling in the systems of computer mathematics into the proper structure of mathematical disciplines. This, in its turn, leads to the necessity of the development of a methodical support of integrated learning of physical-mathematical disciplines on the base of mathematical and computer modeling in the media of computer mathematics: this in turn certainly \textit{changes the paradigm of the physical-mathematical education. }

There arises the question on the third stage of the process of mathematical modeling, i.e., the realization of mathematical model by computer means, which constitutes the computer modeling. As was demonstrated in a series of studies (see \cite{Kap}, \cite{Mat}, \cite{Gol}, \cite{Dyak1}, \cite{BusO}, \cite{Kor}, \cite{Alad}, \cite{Kir}, \cite{IgnMon}  and others), the certain mathematical packages ideally fit for these purposes . The specificity of applying computer mathematical packages to the system of mathematical education is described in the above works quite comprehensively.  However, in these papers, the application mathematical packages were considered as an additional tool used for the intensification of educative process and a tool giving more illustrative features to the studied mathematical structures\footnote{ A propos, similar ideas can be found in research of Canadian authors (Buteau et al., 2014). }.  Therefore, summing up, one can say that \textit{the main idea of implementation of the information technologies to the structure of physical-mathematical education is the computer modeling in systems of computer algebra}. We should note that, essentially, this idea is not completely new: it was implicitly formulated (to one extent or another) in works of different authors\footnote{ One should note also Russian language websites dedicated to the application of CAS to the system of physical-mathematical education: www.exponenta.ru and  vuz.exponenta.ru.} -- however in our work we first formulate this idea explicitly; second, we describe a mechanism of its realization, and third, we populate this mechanism with concrete necessary details.

\section{The Choice Of The Computer Algebra System For The Concept Realization}

Numerous investigation carried out by various authors (see, for example, \cite{Ign3},  \cite{Alad} and others) show that, among known systems of computer mathematics, Maple software could be considered as one of the most appropriate systems of computer mathematics, SCM (CAS) for the purposes of physical-mathematical education due to several reasons which include both the cost, the simplicity of the interface, and the correspondence of programming language to standard mathematical language.  In particular, in the monograph by \cite{Alad}, dedicated to comparative aspect of Maple and Mathematica CAS, the following circumstance was noted: ``CAS Maple supporting quite well-developed procedure programming language, corresponds in best way to problems of educative character and, in particular, to the problems of enhancement of teaching for the mathematics-oriented disciplines in universities, uptake of systems of computer mathematics and application in automation of analytical and numerical  transforms as well as calculation in relatively uncomplicated scientific-technical projects\dots  Maple turned to be simpler in acquisition mainly due to the fact that its language by its syntax is more close to known imperative programming languages, in particular, to Pascal. Next, as is known, in general case, the imperative languages are simpler in adoption''.  The advantages of CAS Maple in comparison with other systems of computer algebra in solving problems of physical-mathematical education are listed also in the monograph by \cite{IgnMon}, dedicated to application of CAS Maple to problems of computer modeling of fundamental objects and phenomena studied in mathematical and physical courses of universities. One of important advantages of CAS Maple for educative system is the excellent quality of 3D dynamical graphics (this fact was revealed even more clearly in the last versions of Maple 17-18\footnote{ The same circumstance is noted by well-known specialist in CAS \cite{Dyak1}. }), as well as simple tools for creation of self-written libraries of procedures.  Taken together, all these features bring CAS Maple to the leading position in the system of physical-mathematical education.

\section{Ideology Of The IT Implmentation}

To implement information technologies into the structure of physical-mathematical education one should solve the following scientific-methodological problems:

\begin{enumerate}
\item  Create the information support of the educative process:
\begin{enumerate}
\item  create electronic manuals;
\item  create generators of individual tasks;
\item  create an automated system of checking the individual tasks;
\item  create electronic libraries.
\end{enumerate}
\item  Create demonstrative accompaniment of lectures and practical/lab works:
\begin{enumerate}
\item  create interactive 3D illustration of geometrical and physical objects;
\item  create interactive video-materials which accompany computations;
\item  create animated mathematical models of objects and phenomena.
\end{enumerate}
\item  Incorporate computer calculations into the structure of practical/lab works:
\begin{enumerate}
\item  Create class-rooms for complex-characrterlessons with application of computers for all physical-mathematical subjects;
\item  incorporate parallel maintenance of practical/lab works by computer calculations;
\item  create programs for analytical testing and students' self-testing.
\end{enumerate}
\item  Incorporate computer calculations into the structure of special courses, annual works, and  graduate qualification works:
\begin{enumerate}
\item  make the construction of computer mathematical models he basement of special courses;
\item  make the creation of author programs and research products as well as interactive study guides by students an obligatory element of graduate qualification works.
\end{enumerate}
\end{enumerate}
 The scheme of educative process depicted in Fig. 2 proposes the solution of the following educative-scientific problems:
\begin{enumerate}
\item \begin{enumerate}
\item  creation of computer models of studied phenomena, attraction of information technologies to the process of teaching a subject;
\item  creation of interactive study guides and systems of analytical testing;
\item   attraction of methods of symbolic mathematics to description of complex phenomena;
\item  replacement of an academic method of teaching subject by an interactive one with application of the information technologies;
\item  usage of  the IT for re-orientation of young people interests to research creativity;
\end{enumerate}
\end{enumerate}

\Fig{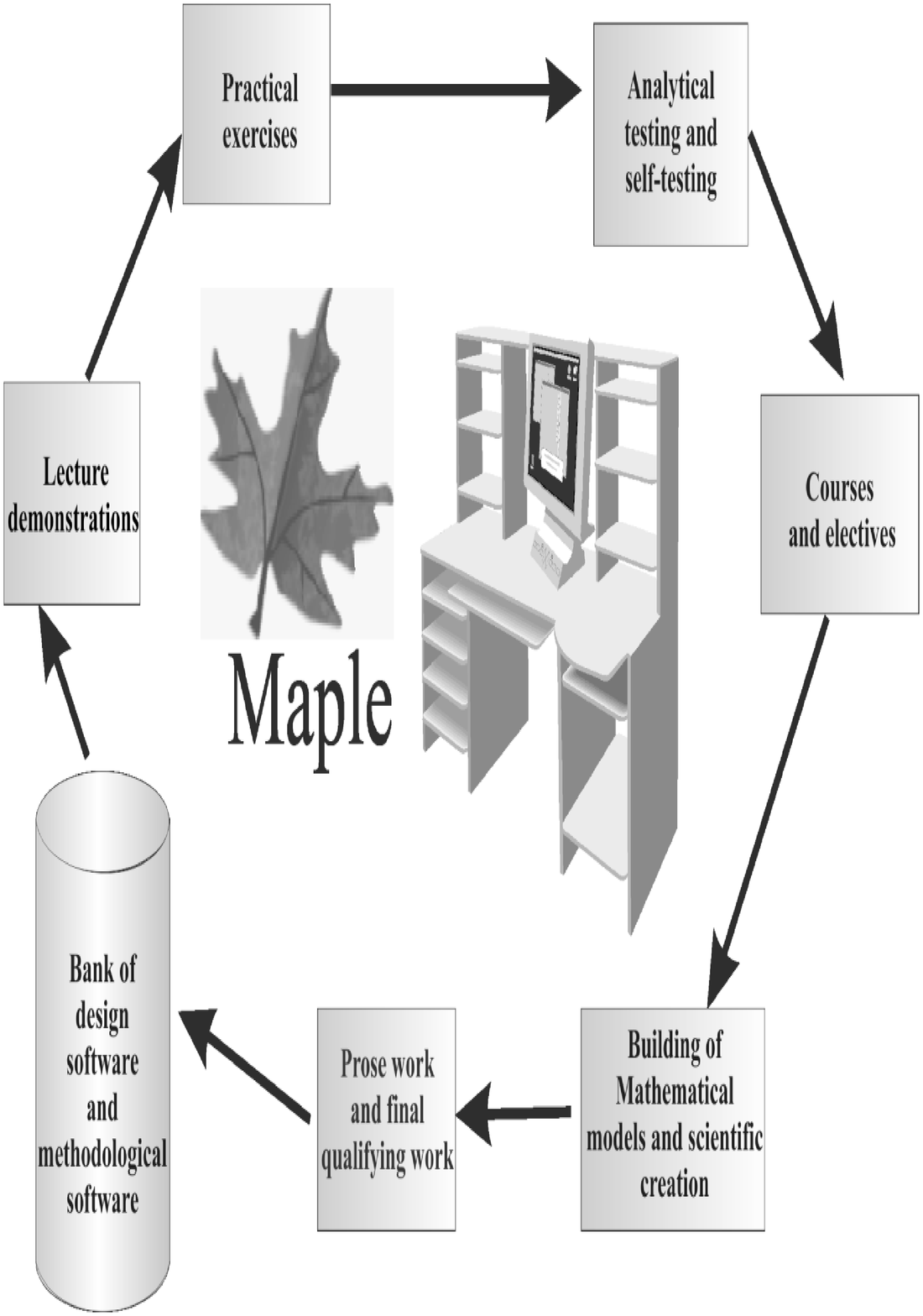}{8}{6}{Organization of educative process for physical-mathematical disciplines on the base of CAS}

\section{
The Methodical And Software Provision Of The IT Implementation Into The Structure Of Physical And Mathematical Education}

Organization of the academic cycle mentioned above with broad application of the information technologies on base of CAS requires large science-based investments  both to the  stage of running up the educative process and to all posterior stages. Even in the first stages of the educative process a large number of developed in advance computer models of studied objects will be required for both demonstration during lectures and use in seminars and student independent works. The computer models developed for educative process must comply with a series of obligatory requirements:
\begin{enumerate}
\item  they must be illustrative;
\item  they must reflect all basic properties of the investigated model;
\item  they must be interactive, i.e., they should allow user to manipulate them with the help of the external devices;
\item  they must be multi-parametrical to ensure the possibility to carry out numerical experiments.
\end{enumerate}

The problem of provision of demonstrativeness of mathematical structures plays an important role in higher education, since the acquisition of fundamental geometrical concepts prepares a base for understanding of the process of mathematical modeling and acquisition of methods of computer modeling. The latter in its turn creates the prerequisites for innovative development of the modern education.  Let us note that the multi-parametric feature of created computer models is the most important factor which makes it possible to control the mathematical model, in other words, to carry out computer modeling.  In connection with the latter, the computer visualization of mathematical models is of importance, especially, the   \textit{equipped dynamic visualization}, basic principles of which were developed in works by \cite{Ign2},\cite{Ign3},\cite{IgnMon}. The creation of complex computer models of that sort is possible in the form of independent software packages (software libraries), which can be used by both educators and students via calling the corresponding libraries and multi-parametric commands contained there and possessing a simple syntax (see, for example, \cite{Ign3}, \cite{Ign4}, \cite{IgnSam},\cite{IgnSam1}). It is necessary to outline that an increase of the demonstrativeness and interactivity levels of educative materials created by means of IT requires investment of large intellectual efforts and high degree of professionalism of teachers creating such programs. Let us note that versions 17--18 of Maple contain the interactive program procedures called `Tutor' in the `Student' library, which allows to display the results into Maplet windows. Based on these, one can create demonstrative and methodical materials. However, the use of these procedures is not enough of course for sufficiently high level of studying the higher mathematics.

The solution of the problem of computer realization of the linear algebra's and analytic geometry objects and creation of demonstrative geometric images (interpretations) of objects, structures, and properties assume the solution of three main tasks:
\begin{enumerate}
\item  construction of mathematical models of basic algebraic structures, objects, and properties;
\item  construction of their geometrical interpretations, i.e., setting the correspondence  of them to the respective geometrical models;
\item  construction of multi-parametrical computer models of objects graphical
\end{enumerate}

Let us note that the multi-parametrical feature of created computer models is the most importance factor of computer models which allows to control the mathematical model, i.e., to carry out computer modeling.  The most efficient solution of these problems is possible in the Systems of Computer Mathematics (SCM), among which Maple system is found to be the most convenient for achieving educative purposes.  The basic advantages of this system with respect to education goals are: relatively low costs (compared with MatLab and Mathematica), user-friendly and interactive interface, perfect graphical features, in particular, interactive 3D- and dynamical graphics (animation).  In this paper we consider the main principles of mathematical and computer modeling of objects of linear algebra and analytic geometry in CAS Maple. Note that for program procedures considered here the version of Maple (starting with ver. 6) is not important.

It is necessary to underline that an increase of the degree of demonstrativeness and interactivity of educative materials created by means of IT requires an input of greater intellectual efforts and higher degree of educators' professionalism. First of all, this concerns the subjects of physical-mathematical cycle. Here the central idea of creation of high-quality electronic educative materials consists in mathematical modeling of objects and phenomena studied. Creation of a mathematical model of the studied object determines, in many aspects, the demonstrativeness and the degree of acquiring of the subject learned. Therefore the basic requirements with respect to the mathematical model must be: its multi-parametrical feature, possibility of 3D-graphical realization, interactivity, possibility to construct animated (graphical dynamic) representations.

The systems of computer mathematics (CAS), and Maple first of all, represent uniquely exceptional program and graphic capabilities for realization of this idea \cite{BusV}). However, an attempt to apply directly standard procedure of CAS not always gives us the desired result. To obtain quality graphic and animated models of basic mathematical structures of function analysis, one should create user multi-parametrical software procedures, which were simple for a user non-experienced in programming and were convenient to join into specialized libraries of user procedures \cite{Dyak1}).

In what follows we give an example of creation and application of such a library in the theory of curves of the 2nd order. The theory of curves (surfaces) of the 2nd order, closely related to the theory of quadrics in Euclidean space and theory of their reduction to canonical form by means of motion transforms, is known to find numerous applications in mathematical analysis, mechanics, and the field theory.  On the other hand, the theory of curves (surfaces) of the 2nd order fails to be acquired sufficiently by students of natural-science specialization and even by students of mathematical divisions. An increase of the quality of acquisition of abstract mathematical matters is possible through integration of methods of mathematical modeling and equipped dynamic visualization in the computer mathematical package Maple (\cite{Dyak1}, \cite{Alad}, \cite{BusV}). In Russian universities, a sufficiently complete theory of curves and surfaces of the 2nd order is given, containing, along with general classification of curves and description of their main properties and elements, also the problems on reduction of a general equation of the 2nd order in plane (in space) to the canonical form. In order to solve these problems we developed a package of programs for automatic complete investigation of general equation of the 2nd order in plane with the output of the results of investigation in both table form (containing the information on curve type, formula of transformation of equation of the 2nd order to canonical form, curve's parameters and all its canonical elements, canonical equation of the curve) and as the respective curve's graphics with all its elements depicted along with the initial and canonical systems of coordinates. In addition, the program determines automatically the optimal parameters for curve depicting and domain of its depiction. The package differs from known application mathematical packages by, firstly, simplicity of program input which is given by a single command with only a help of general equation in plane and user parameters for output of the results of investigation. Secondly, it provides user with complete representation of investigation results in textual, analytic, and graphic forms. These results are displayed in form of  matrix equipped with necessary explanatory text determined by the results of curve investigation.  The investigation of a curve is carried out by means of three-parametrical procedure AnalGeo CanonF](Eq,X,X1,s), where Eq is the general equation of the curve of the 2nd order, X is a list of coordinates in the initial system of coordinates in format [x,y], X1 is a list of coordinates in new system of coordinates in format [x1,y1], s stands for the name of variable of the systems of coordinates' rotation angle.  At the execution of command, the type of curve is an output as well as matrix of its parameters  (list of eigenvalues of the quadratic form, the canonical equation of curve, motion transformation which leads to the canonical equation, the list of parameters [с,$\epsilon$,d], i.e., [distance from center to foci, eccentricity, distance from center to directrices], [a,b,p] are [values of semi-axes and parameter in the canonical equation of parabola]).  In Fig. 3, 4 the execution of command is shown.

From the figures given above one can see extraordinary features of this software package: a single and simple command runs complete investigation of an arbitrary equation of the 3nd order in plane. In addition, the results are presented simultaneously in two forms: I) the analytic one given as a matrix determining: all elements of canonical form of the quadric, canonical motion, type of geometrical figure obtained, and numerical values of all its parameters, II) the picture of the figure obtained with all its elements depicted and also the initial and canonical systems of coordinates.

\TwoFigs{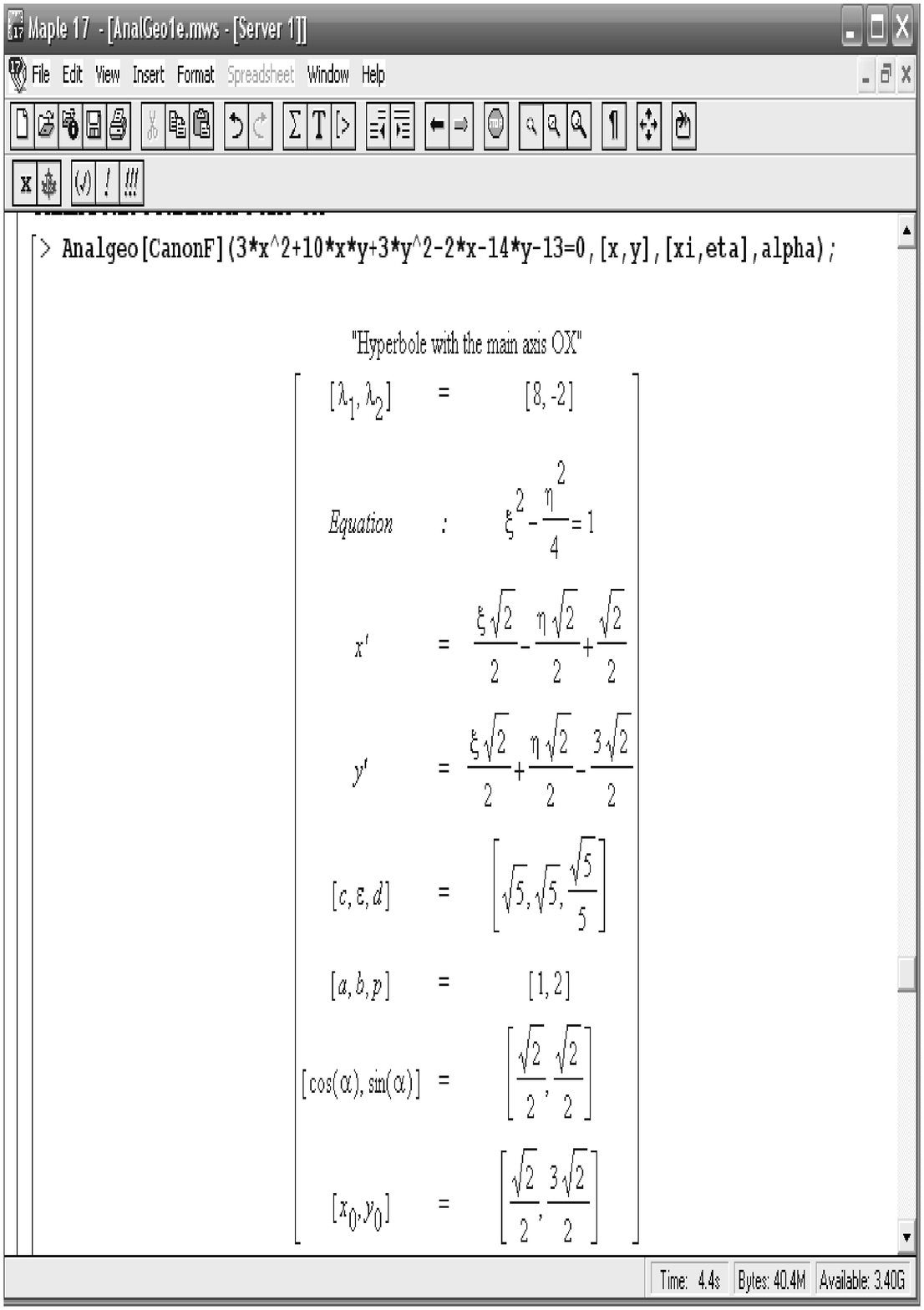}{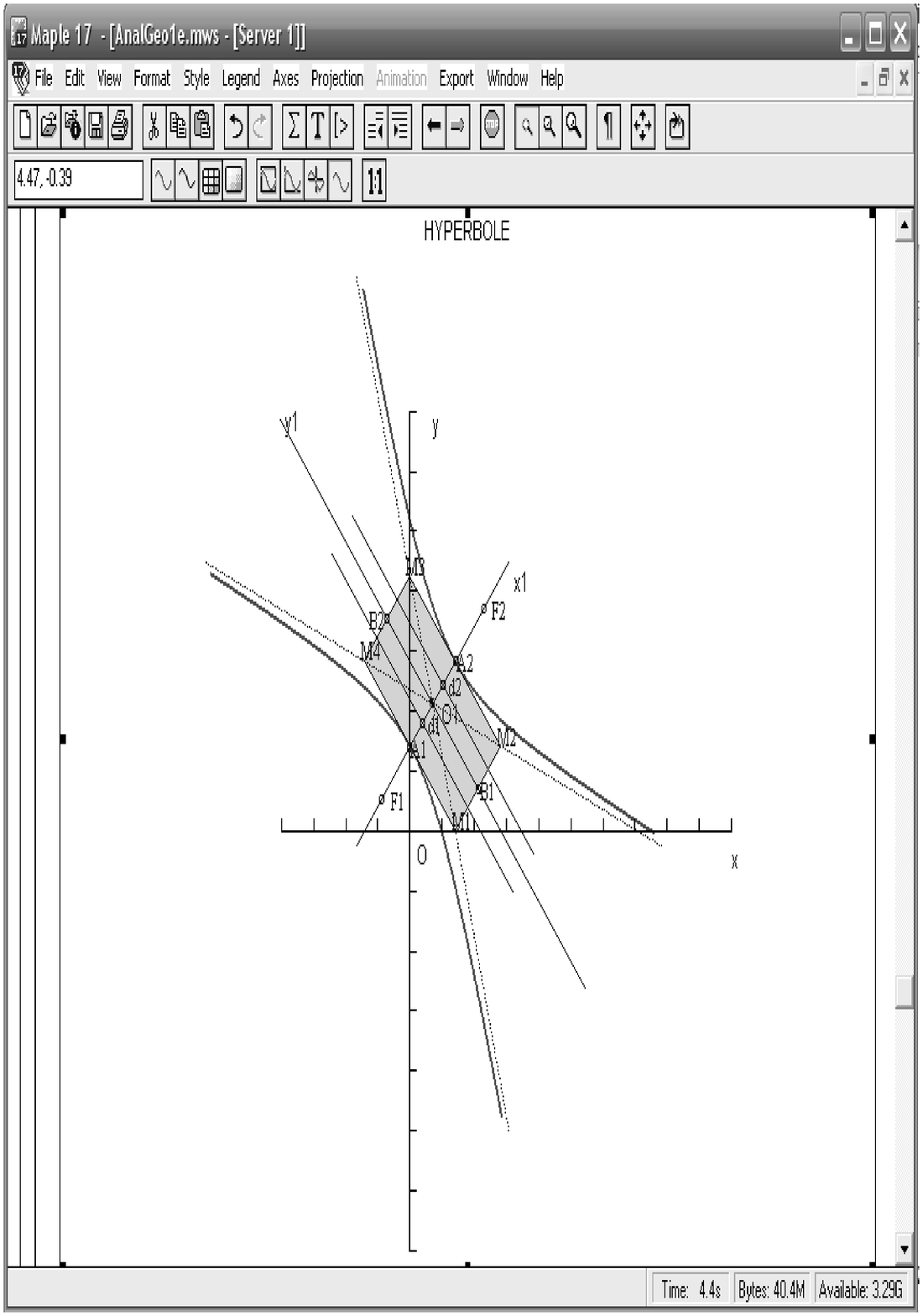}{Output investigation results of the 2nd order equation:
 $3x^2+10xy+3y^2-2x-14y-13=0$ in form of matrix}{Output of the results of investigation of the 2nd order equation $x^2+y^2+4xy-8x-4y+1=0$ in form of picture}{6}

\section{
A Technology Of The Combined Study Of Mathematics And Computer Modeling}

The basic methods of realization of the idea of informatization of subjects of physical-mathematical cycle on the base of mathematical and computer modeling in systems of computer mathematics are the following ones:
\begin{enumerate}
\item  profile orientation of mathematics courses;
\item  application of mathematical modeling method as a base method for learning special subjects;
\item  alignment of the whole system of specialists' preparation around a solution of scientific-technical problems and preparation of diploma project;
\item  building-in of computer modeling in all special courses;
\item  organization of academic activities in special subjects in the form of complex lab research works with application of computer mathematics and IT;
\item  the main criterion for graduation should be a qualification work with mandatory application of computer modeling methods and with either possibility of scientific publishing or direct use of the result in educative process.
\end{enumerate}

The following items are necessary organizational measures for ensuring implementation of information technologies into the structure of physical-mathematical education:
\begin{enumerate}
\item  revise curricular programs of special subjects;
\item  create educative-methodical provision of special courses;
\item  organize respective retraining of educators in domain of computer modeling and IT;
\item  equip modern computer IT labs;
\item  provide these labs with licensed packages of Mathematica, Maple, MatLab, CorelDraw, Delfi, WinEdt, Microsoft Office and other software;
\item  reequip classrooms for seminar lessons into classes for complex learning with a usage of computers;
\item  organize system of summer scientific schools for students and post-graduates in mathematical and computer modeling.
\end{enumerate}

 It should be noted that, in the Institute of Mathematics and Mechanics of Kazan Federal University, these measures are undertaken according to plan in both material support and educative-scientific provision. In what follows, we will show some examples of implementation of IT on the base of methods of mathematical and computer modeling in CAS Maple into system of teaching the higher mathematics.

In Fig. 5 the scheme of organization of classroom for combined learning of mathematics and computer modeling is given, while in Fig. 6 we depict a block-module (workplace) of a student in this classroom, allowing the alumni to work with a same comfort with PC, textbooks, and manuals.

\TwoFigs{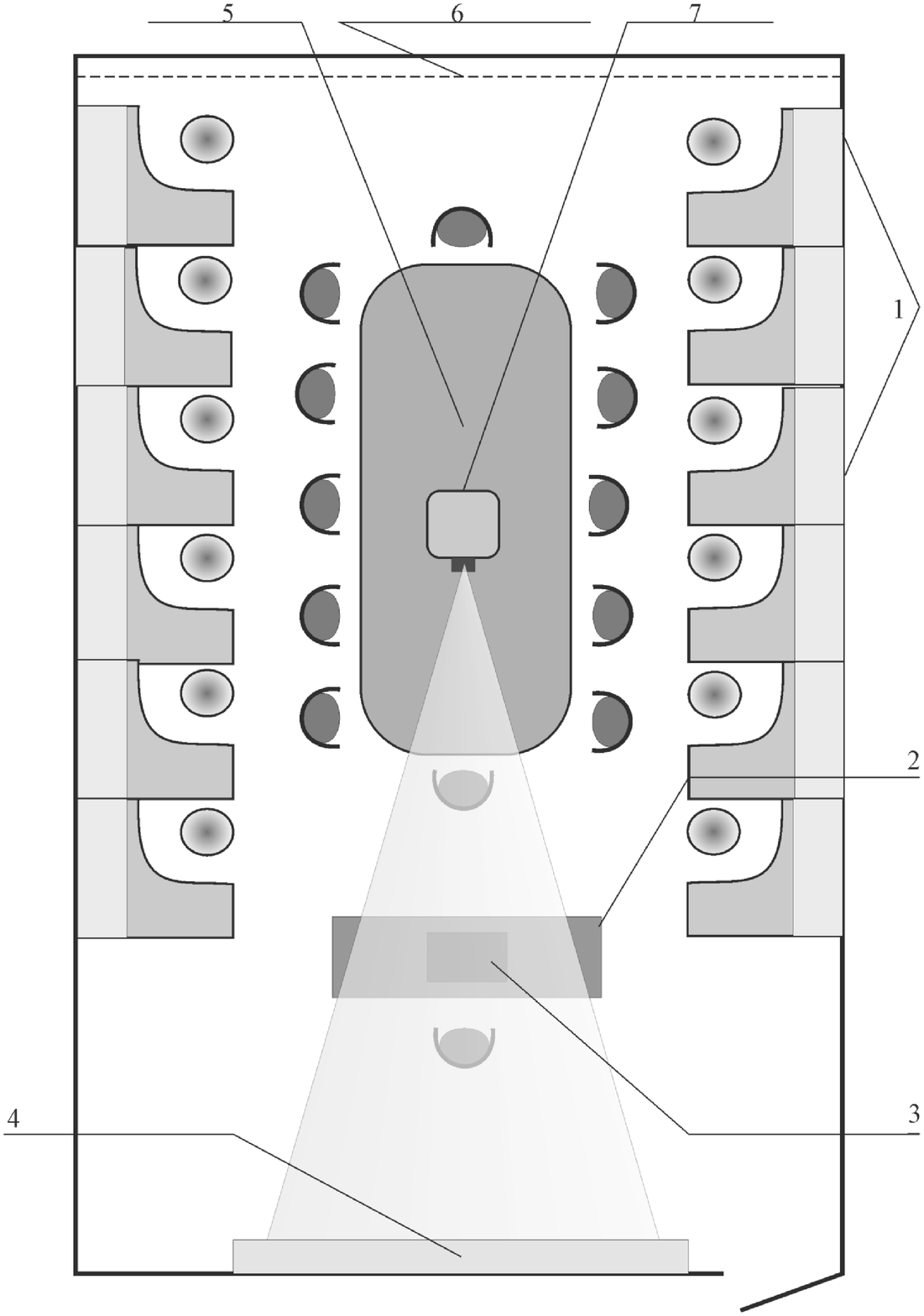}{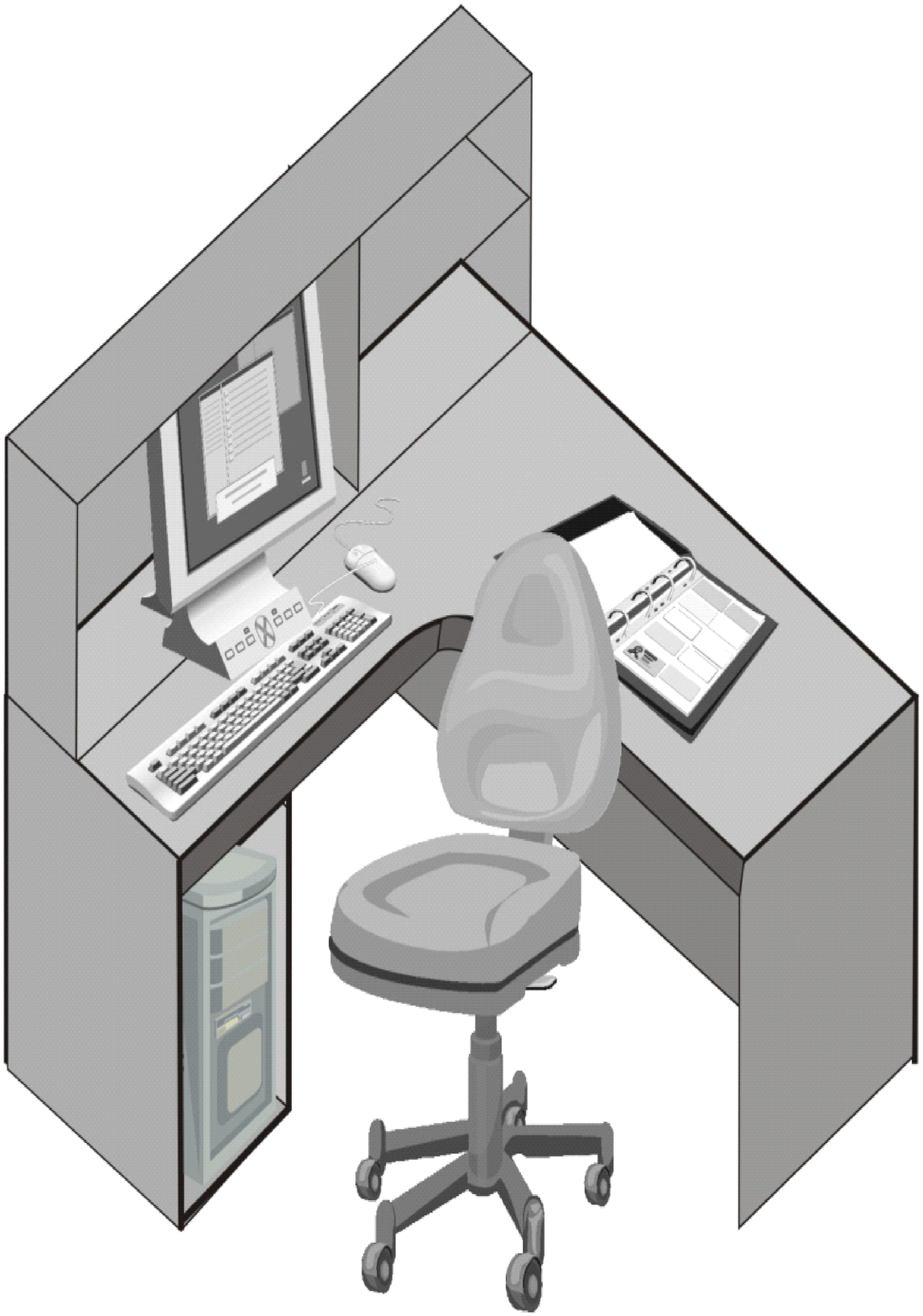}{Computerized classroom for complex learning of subjects of physical-mathematical cycle: 1 modules; 2 educator's table;
  3 educator's PC; 4 interactive whiteboard; 5 discussion table;  6 blinds; 7 projector.}{Student workplace (module) for complex learning with a computer usage}{8}

\section{
A Realization Of An Idea Of Mathematical And Computer  Modeling Methods Implementation To The Process Of Mathematical Training For Teachers In Maple Package}

The software package described above and other similar are integrated with software packages in CAS Maple containing individual tasks for students and providing the tools for the automated analytical (formula) testing and allow students to  study independently the materials and carry out the solution of problems.

For  more complete informatization of mathematical education the idea of \textit{analytical testing }seems to be productive. This idea was first stated by \cite{Ign1} (see also \cite{IgnAd}) and consists of testing the mathematical knowledge of student by means of CAS Maple. Herewith conditions of problem and answer in form of formula are displayed by students in the Maplet window. Here the ability of Maple to compare equivalence of two different expressions, which possibly are exposed in different formats, is used. In contrast to a usual mechanical testing where one must select a right answer from a list (this way one cannot elucidate existing mathematical skills), the analytical testing is substantially an intellectual testing. Thus, it becomes possible to enclose the process of studying mathematics with a verification of the knowledge obtained.
For the final closure of the process of learning higher mathematics on base of CAS Maple, it is necessary to automate the process of knowledge control both in its intermediary and basic levels in form of an exam/final testing and implement these procedures into point rating system of university/institute. To this aim we developed a complex of SRS programs in CAS Maple, which realizes exchange of data from the window of Maplet with working sheets of Microsoft Excel, where to each academic group there corresponds a separated sheet in tables from which write to /read from CAS Maple can be performed. The created program for exam testing has the following structure (Fig. 7):
Block 1 (GenLib.mw) generates a library of program procedures (BRS) for calculating the score of points and presentation of results of exam session in form of a histogram (file Lib.m) (block 2). Block 3 uses the library Lib.m by means of which the Maplet (block 4) is created. User calls Maplet -- program (block 5), which opens the test window (block 7). During the registration for being admitted to exam, Maplet calls to the sheet of Microsoft Excel with results of educative modules passing during the respective semester (block 6а, sheet of Microsoft Excel).  During registration, a window opens, containing the list of student group with information of access to the exam for each of the education modules.  Having chosen a concrete student, the Maplet window is opened (block 9), where educator/examiner inserts marks in five-point scale for the answer to each of the of examination card's questions. At pressing ``RATING'' button, the final results of exam with the results of points gained over semester modules taken into account are displayed in a special window in five-point system, being simultaneously recorded in the respective table of Microsoft Excel sheet (block 6b). At the same time, the nest window of Maplet opens with the result of exam session for this group (block 10), where, after input of group number and pressing ''RATING'' button, in the Maplet window opened (block 11) the results of session in academic group are displayed in both table shape and as a histogram (the histogram is depicted in the figure).
 
\Fig{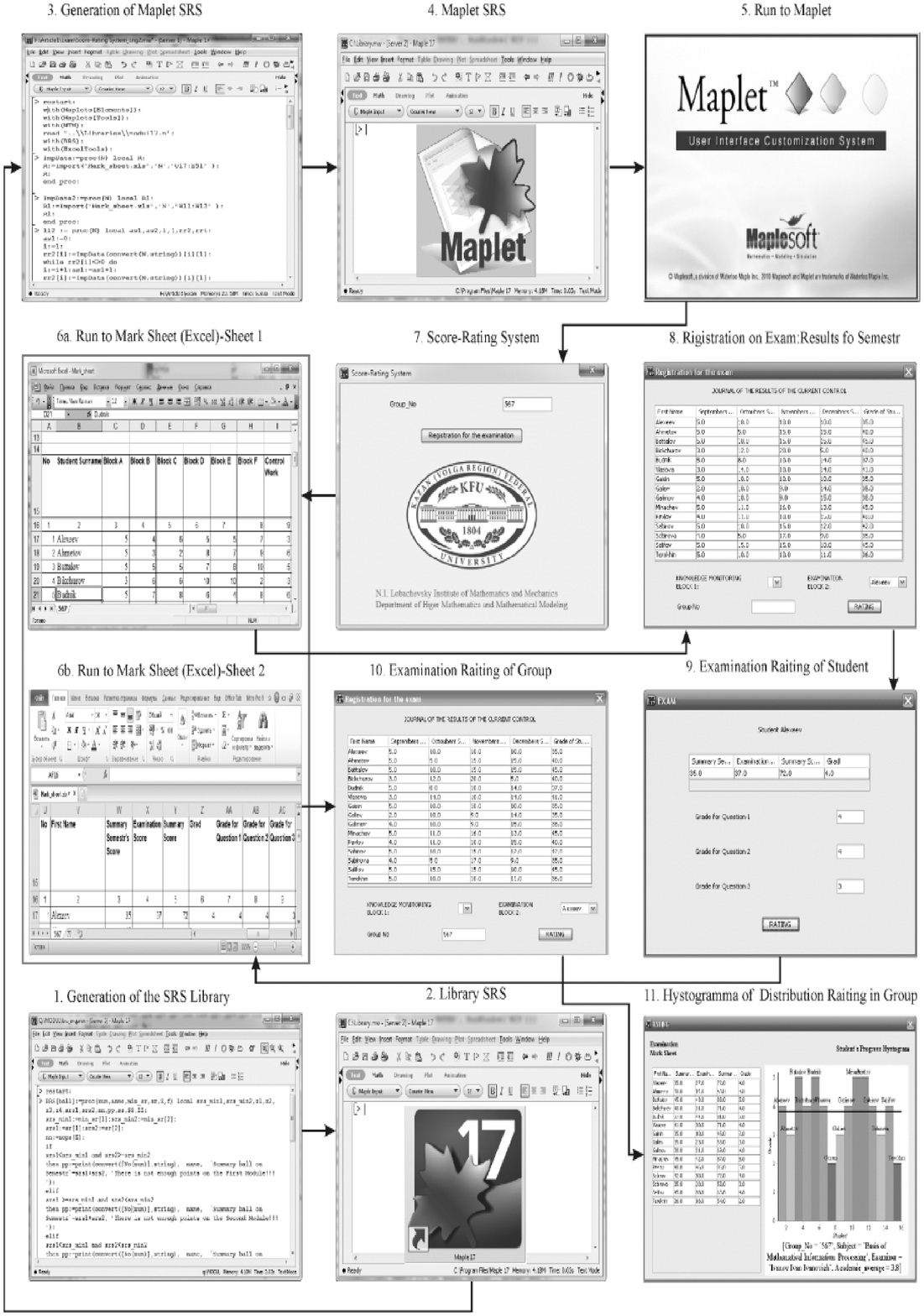}{14}{14}{Flow chart of the program of exam testing SRS.}
 
The programs and IT methods described above were introduced into the process of teaching higher mathematics in 1st and 2nd year academic groups of both Institute of Physical Culture, Sport, Regenerative Medicine and Law Faculty of Kazan Federal University (KFU). Let us note that students of the cited profiles have very low motivation for studying higher mathematics, which they usually treat as a subject not necessary for their future professional activities. For carrying out the experiment on implementing IT on the base of Maple pack, pairs of experimental and reference groups were chosen. In what follows some results of the final control of knowledge and students' testing are presented (see Fig. 8 to 11).

\Fig{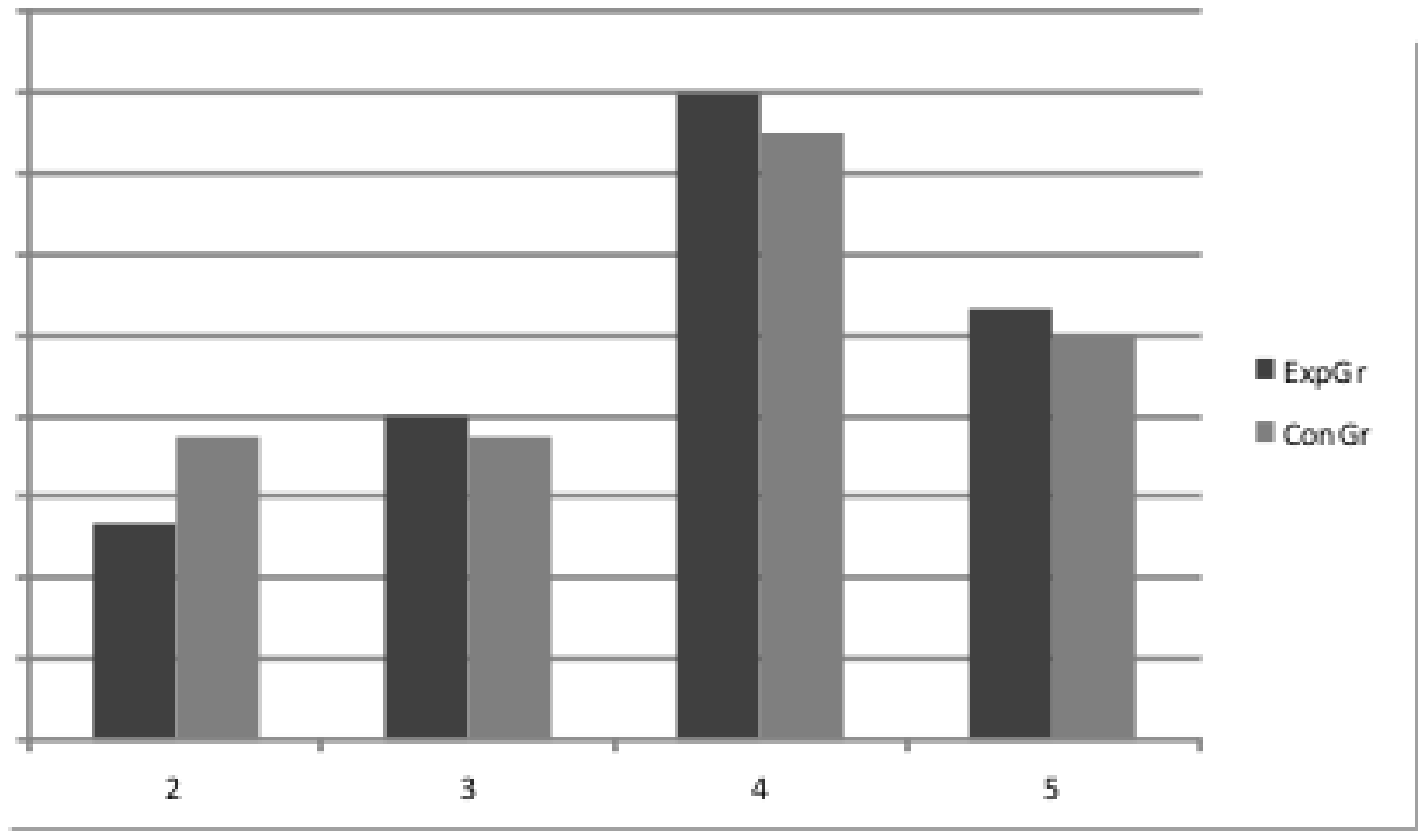}{10}{6}{The results of exam session in experimental and reference groups. The horizontal axis shows marks in five-point system; on the vertical axis the fraction of students received the respective mark is given (percent).}

The analysis of the results represented in Fig. 8 makes it possible to arrive to the following conclusions:
1.  Academic performance in experimental group is above the same in reference group by 5\%;

2.   The quality of knowledge in experimental group exceeds the same in reference group by 9\%.
 In Fig. 9 and 10 the results of questionnaire survey on motivation of learning via CAS are shown after conclusion of the course of higher mathematics.
\TwoFigs{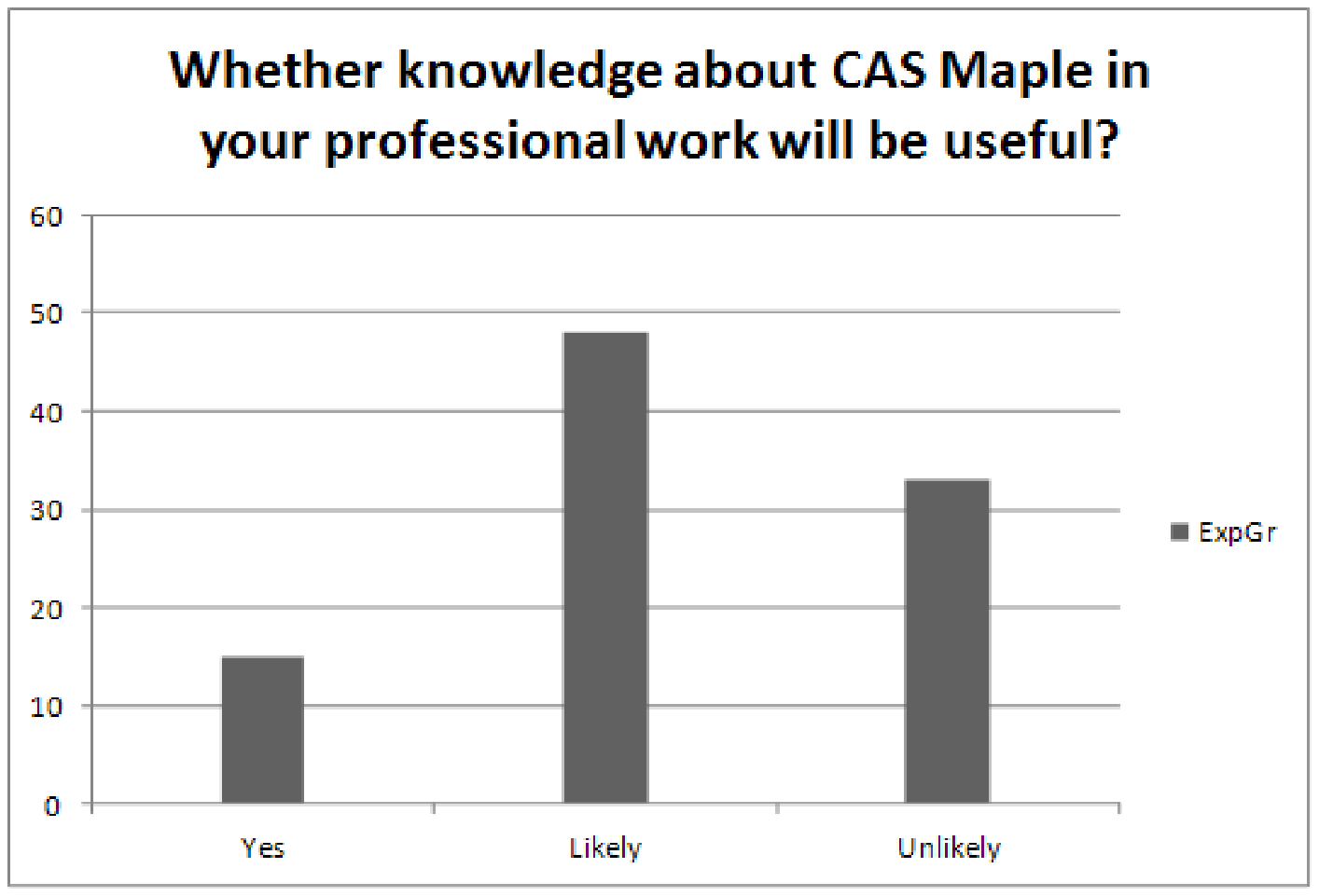}{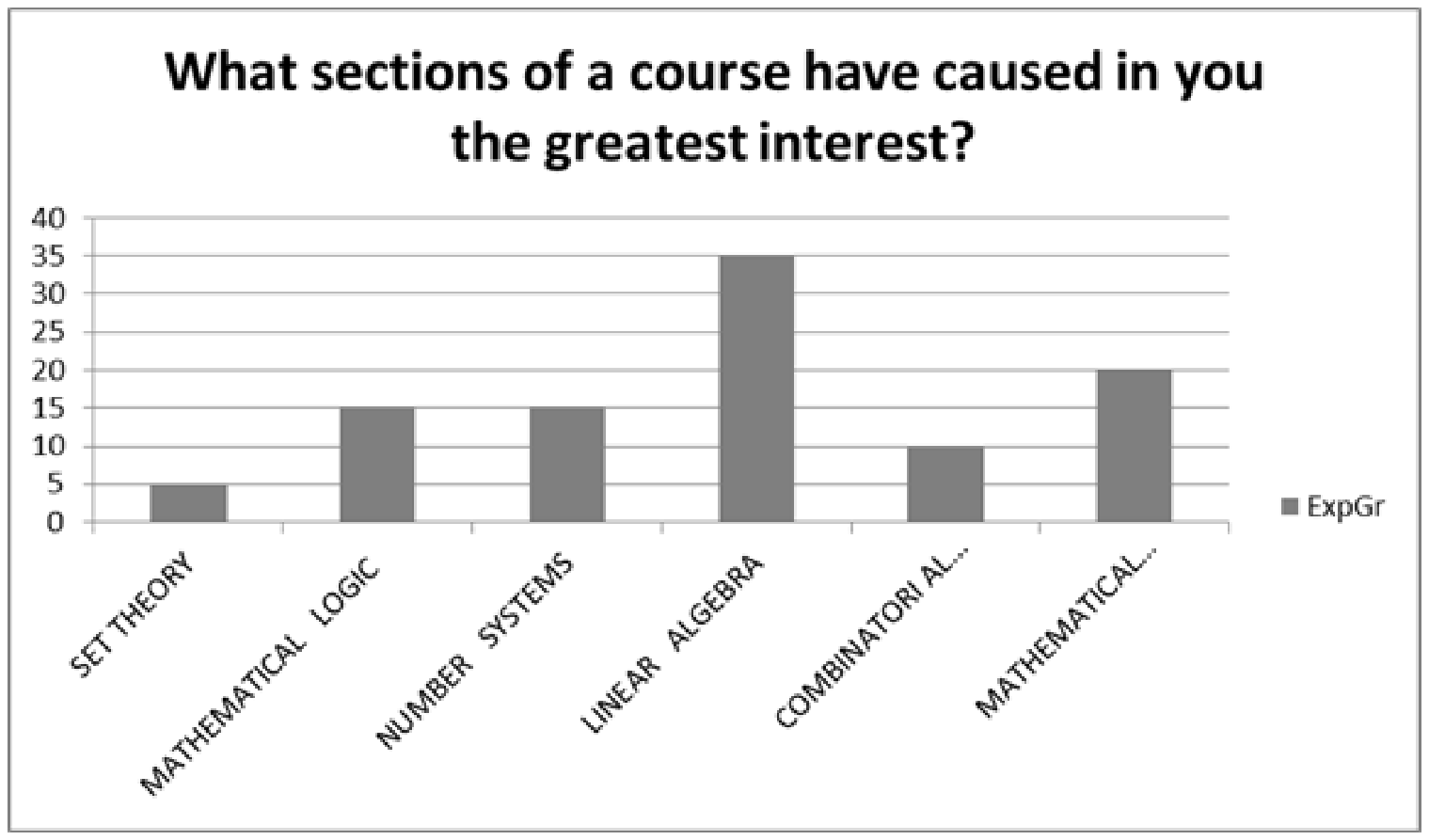}{The results of questionnaire survey in experimental group}{The results of questionnaire survey in the experimental group}{6}

In Fig. 11 the results of questionnaire survey among students of the experimental group during their self-attestation (as concerns their acquisition level in higher mathematics) are depicted.

\Fig{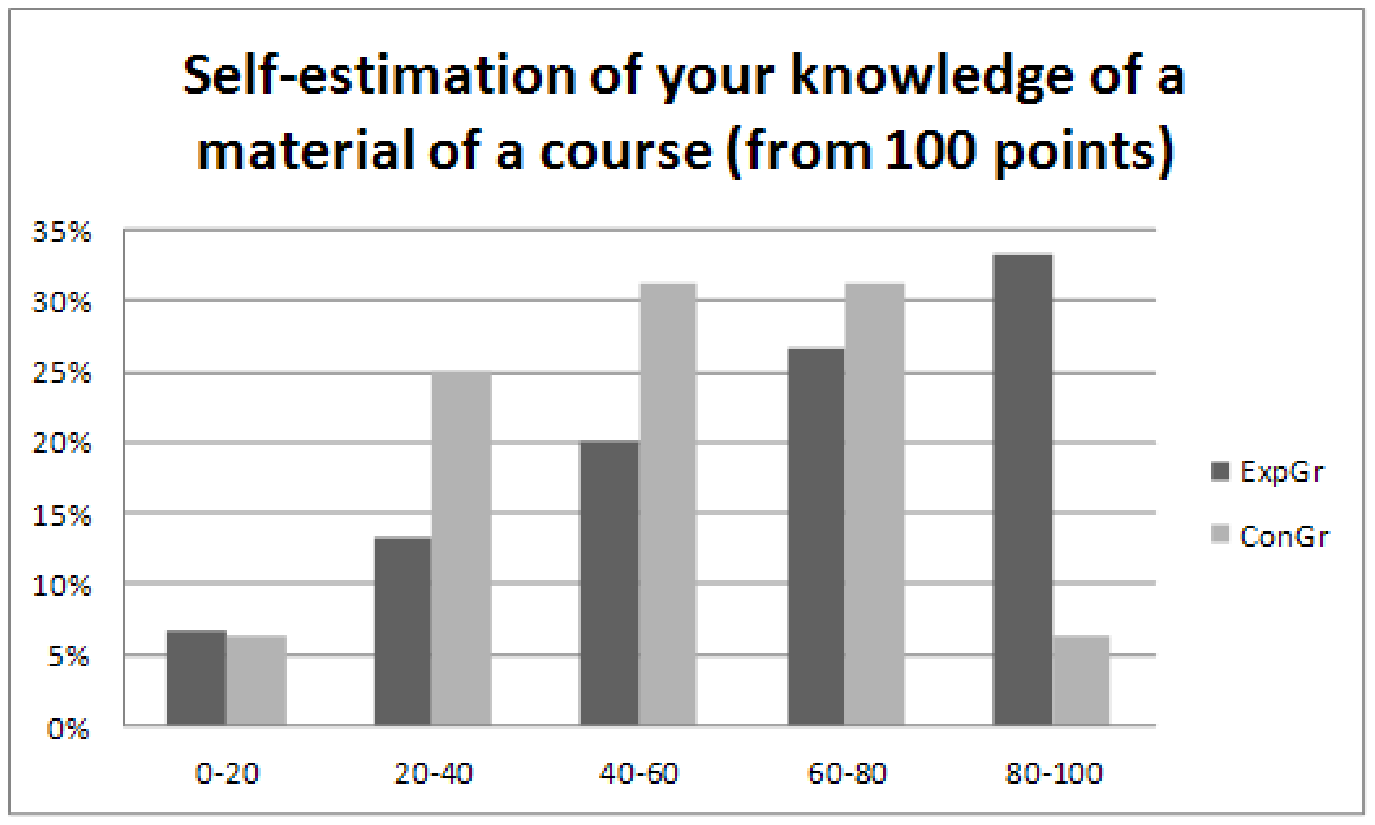}{10}{6}{The histogram of students' self-attestation in experimental and reference groups.  The horizontal axis shows points of self-attestation, the vertical one shows the quantity of students in percentage.}

As one can see from the above figures, a stable trend to the increase of motivation in experimental group to study higher mathematics is visible as well as a higher self-estimation of their knowledge of the subject.

 \section*{The Conclusion}

The experience of implementation of the IT on base of the combined application of methods of mathematical and computer modeling in CAS Maple into system of mathematical education in Kazan University thus led to positive results in both academic progress of students and an increased interest of non-mathematical specialties' students to learning higher mathematics. In addition, it should be noted that student from experimental group gained simultaneously additional competences in information technologies, which undoubtedly will serve to their professional growth in the future.

 \section*{The Acknowledgment}

This work was funded by the subsidy allocated to Kazan Federal University for the state assignment in the sphere of scientific activities. The authors express their gratitude to the administrative office of Kazan University and directorate of N.I. Lobachevsky Institute of Mathematics and Mechanics for material support in necessary hardware and software and other equipment as well as licenses. In addition, we would like to extent our gratitudes to Dr. Alexander Agathonov for valuable help in the development of testing program codes.


\begin{thebibliography}{99}

\bibitem{IgnAd}%
Adiyatullina, G.R., Ignat'ev, Yu. G. (2011). The Interaction of Maplet with date Base in formats txt and xsl in the Analytic Systems of Testing. \textit{Vestnik TGGPU}, CityplaceKazan, country-regionRussia, 3(25),  21-25.

\bibitem{Alad}%
Aladjev, V.Z., Bojko, V.K., Rovba, E.A. (2011). Programming in packages Maple and Mathematica: Comparative aspect. CityGrodno (country-regionplaceBelarus), Noosphere Publ. (In Russian).

 \bibitem{Bute}%
 Buteau, C., Jarvis, D. H.,  Lavicza, Z. (2014).  On the Integration of Computer Algebra Systems (CAS) by Canadian Mathematicians: Results of a National Survey. \textit{Canadian Journal of Science, Mathematics and Technology Education}, 14:1, 35-57, DOI:10.1080/14926156.2014.874614

\bibitem{BusV}%
 Bushkova, V.A. (2011).  Library of program procedures of creation of the operated clothing dynamic visualization of geodesic lines in CAS Maple. \textit{Vestnik TGGPU}, CityplaceKazan, country-regionRussia, 26, 8--10.

\bibitem{BusO}%
Bushkova, O.A. (2006). Design  of  Computer Geometry Resource in ``Mathematica'' Environment.  Open Education. No 6, 18-22.  (In Russian).

\bibitem{Dyak1}%
 Dyakonov, V.P. (2001). Computer Mathematics. \textit{Soros Educational Journal, }1, 116--121.

\bibitem{Dyak2}%
 Dyakonov, V.P. (2006). Maple 9.5/10 in Mathematics, Physics and Education. placeCityMoscow, SOLON-PRESS. (In Russian).

\bibitem{Gol}%
 Goloskokov, D.P. (2004). The Equations of Mathematical Physics: Solving problems in the Maple System. St. Peterburg, Piter. (In Russian).

\bibitem{Ign1}%
 Ignat'ev, Yu.G. (2005). Problems of IT in Mathematical Education. Kazan, TSHPU Publ.

\bibitem{Ign2}%
 Ignat'ev, Yu.G. (2009). The user graphic procedures for creation of animated models of nonlinear physical processes. \textit{Computer algebra systems and their applications: Proceedings of the International Conference. }CityplaceSmolensk, SmolGU, 10, 43-44. (In Russian).

\bibitem{Ign3}%
Ignat'ev, Yu. G., Abdulla, K. H. (2010a). Mathematical Modeling of Nonlinear Generic Mechanics System in Computer Mathematics Maple. \textit{Vestnik PFUR: Mathematics, Informatics and Physics.} No 4(3), 99-111. (In Russian).

\bibitem{Ign4}%
Ignat'ev, Yu. G., Abdulla, K. H. (2010a). Mathematica; Modeling of Nonlinear Electrodynamics Systems  in Computer Mathematics Maple. \textit{Vestnik TGHPU: Physical and Mathematical Sciences.} 2(20), 22-27 (In Russian).

 \bibitem{IgnSam}%
 Ignat'ev, Yu. G., Samigullina A.R. (2011a). The Library of Program Procedures for Methodical Support of Higher Algebra Course in the Computer algebra system Maple. \textit{Vestnik TGHPU: Physical and Mathematical Sciences.} 1(23), 21-24 (In Russian).

\bibitem{IgnSam1}%
 Ignat'ev, Yu. G., Samigullina A.R. (2011b). The Software Support of the Theory of Curves of Second Order in Computer Mathematics. \textit{Vestnik TGHPU: Physical and Mathematical Sciences.} 4(26), 24-29 (In Russian).

\bibitem{IgnSam2}%
 Ignat'ev, Yu. G., Samigullina A.R. (2011). The Program of Exactl Calculation of Fundamental Solutions of Systems of Algebraic Linear Equations and Representation of Them in Standard List Form in Mathematical Package Maple. \textit{Bull. OBPBT, No 3(76), 547.}

 \bibitem{IgnMon}%
 Ignatyev, Yu. G. (2014). Mathematical modeling of fundamental objects and phenomena in system of computer mathematics Maple. Lectures for school on mathematical modeling. CityKazan, PlaceNameplaceKazan PlaceTypeUniversity Publ. (In Russian).

\bibitem{Kap}%
 Kapustina, T.V. (1999). Computer System Mathematica 3.0 for Users. placeCityMoscow, SOLON-PRESS. (In Russian).

\bibitem{Kir}%
 Kirsanov, M.N. (2012). Maple  and Maplet. Solving Mechanics Problems. St. Peterburg, Lan. (In Russian).

\bibitem{Kor}%
 Kornilov, V.S. (2007). Modern information and communication technologies in humanitarian studies of mathematical models of inverse problems for differential equations. Vestnik PFUR: Informatization of Education. No 1,  64-98. (In Russian).

\bibitem{Mat}%
Matrosov, A.V. (2001). Maple 6. Solving Problems of Higher Mathematics and Mechanics. St. Peterburg, BHV-Peterburg. (In Russian).

\bibitem{Samar}%
 Samarsky, A.A. , Tikhonov A.P. (2005). Mathematical Modelling:  Ideas. Methods. Samples. \textit{Moskow, Fizmatlit.} (In Russian).

 \end{thebibliography}
\end{document}